\documentclass[%
reprint,
superscriptaddress,
amsmath,amssymb,
aps,
]{revtex4-2}

\usepackage{graphicx}
\usepackage{dcolumn}
\usepackage{bm}

\usepackage[T1]{fontenc}
\usepackage[utf8]{inputenc}
\usepackage[normalem]{ulem}

\usepackage[dvipsnames]{xcolor} 
\usepackage{longtable}  

\begin{document}
	
	\preprint{APS/123-QED}
	
	\title{\emph{Ab initio} study of the structural, vibrational and optical properties of potential parent structures of nitrogen-doped lutetium hydride}
	
	\author{{\DJ}or{\dj}e Dangi{\'c}}
	\email{dorde.dangic@ehu.es}
	\affiliation{Fisika Aplikatua Saila, Gipuzkoako Ingeniaritza Eskola, University of the Basque Country (UPV/EHU), 
		Europa Plaza 1, 20018 Donostia/San Sebasti{\'a}n, Spain}
	\affiliation{Centro de F{\'i}sica de Materiales (CSIC-UPV/EHU), 
		Manuel de Lardizabal Pasealekua 5, 20018 Donostia/San Sebasti{\'a}n, Spain}	
	\author{Peio Garcia-Goiricelaya}
	\affiliation{Centro de F{\'i}sica de Materiales (CSIC-UPV/EHU), 
		Manuel de Lardizabal Pasealekua 5, 20018 Donostia/San Sebasti{\'a}n, Spain}	
	\author{Yue-Wen Fang}
	\affiliation{Fisika Aplikatua Saila, Gipuzkoako Ingeniaritza Eskola, University of the Basque Country (UPV/EHU), 
		Europa Plaza 1, 20018 Donostia/San Sebasti{\'a}n, Spain}
	\affiliation{Centro de F{\'i}sica de Materiales (CSIC-UPV/EHU), 
		Manuel de Lardizabal Pasealekua 5, 20018 Donostia/San Sebasti{\'a}n, Spain}	
	\author{Julen Iba\~{n}ez-Azpiroz}
	\affiliation{Centro de F{\'i}sica de Materiales (CSIC-UPV/EHU), 
		Manuel de Lardizabal Pasealekua 5, 20018 Donostia/San Sebasti{\'a}n, Spain}	
    \affiliation{IKERBASQUE Basque Foundation for Science, 48013 Bilbao, Spain}
	\author{Ion Errea}%
	\affiliation{Fisika Aplikatua Saila, Gipuzkoako Ingeniaritza Eskola, University of the Basque Country (UPV/EHU), 
		Europa Plaza 1, 20018 Donostia/San Sebasti{\'a}n, Spain}
	\affiliation{Centro de F{\'i}sica de Materiales (CSIC-UPV/EHU),
		Manuel de Lardizabal Pasealekua 5, 20018 Donostia/San Sebasti{\'a}n, Spain}
	\affiliation{Donostia International Physics Center (DIPC),
		Manuel de Lardizabal Pasealekua 4, 20018 Donostia/San Sebasti{\'a}n, Spain}
	
	
	\date{\today}
	
	\begin{abstract}
 The recent report of near-ambient conditions superconductivity in a nitrogen-doped lutetium hydride has inspired a large number of experimental studies with contradictory results. We model from first principles the physical properties of the possible parent structures of the reported superconductor, LuH$_2$ and LuH$_3$. We show that only the phonon band structure of LuH$_3$ can explain the reported Raman spectra due to the presence of hydrogens at the interstitial octahedral sites. However, this structure is stabilized by anharmonicity only above 6 GPa. We find that the intriguing color change with pressure in the reported superconductor is consistent with the optical properties of LuH$_2$, which are determined by the presence of an undamped interband plasmon. The plasmon blue-shifts with pressure and modifies the color of the sample without requiring any structural phase transition. Our findings suggest that the main component in the experiments is LuH$_2$ with some extra hydrogen atoms at octahedral sites. None of LuH$_2$ and LuH$_3$ superconduct at high temperatures.
	\end{abstract}
	
	\maketitle
	
	\emph{Introduction.}
    One of the most important problems of modern physics is finding a superconductor with a high critical temperature~\cite{problems}. In the last ten years, high-pressure hydrides have emerged as promising candidates to achieve this goal~\cite{review1,review2}. After the initial confirmation of high-temperature superconductivity in H$_3$S at 140 GPa~\cite{H3S}, a large number of different materials have been found to superconduct above liquid nitrogen temperature~\cite{LaH10_1,LaH10_2,YH6,YH9,CaH6}. However, all of these materials are only stable at high pressures and thus not technologically relevant. This instigated a secondary goal, finding a superconducting hydride at a near ambient pressure.  
	
	Dasenbrock-Gammon et al. have recently reported a near-ambient pressure room temperature superconductor in a nitrogen-doped lutetium hydride~\cite{Dias}. On the basis of Raman scattering, X-ray diffraction (XRD), and energy-dispersive X-ray experiments, the material has been identified with the  $Fm\bar{3}m$ space group and the LuH$_{3-\delta}$N$_{\epsilon}$ stoichiometry, where  the possibility of both N-substitution (with concentration of $\epsilon$ per Lu atom) and H-vacancy defects (with concentration of $\delta$ per Lu atom) are remarked. The material has been reported to have a maximum superconducting critical temperature of 294 K at 1 GPa, accompanied by abrupt changes in the magnetic susceptibility, heat capacity, and resistance of the material. The paper reports as well an unexpected change in the color of the material with increasing pressure from blue to pink at 0.3 GPa, and finally to red at 3 GPa. The color change is assigned to a phase transition. Finally, the Raman spectra reveal prominent phonon modes at three clearly separated energy ranges: below 150 cm$^{-1}$, around 250 cm$^{-1}$, and around 1200 cm$^{-1}$.
	
	This finding has inspired a large number of experimental and theoretical works~\cite{LuH3_SSCHA, Colorchangetheory, monserrat,Color_change_LuH2,LuH_color_change_higher_pressure,Shan_2023,zhang2023pressure,luh2_raman,LuH3_ramman}. Unfortunately, none of these works has been able to reproduce the high-temperature superconductivity. Additionally, while qualitatively agreeing, most experiments quantitatively differ from each other. Firstly, the work in Ref.~\cite{Color_change_LuH2} reports the same sequence of colors of the material under pressure as in Ref.~\cite{Dias}, but in pure LuH$_2$. However, the  pressures at which the color of the sample changes are different, as well as for different samples investigated, implying a large inhomogeneity of the samples. The color change has been reported in a number of different works as well~\cite{LuH_color_change_higher_pressure, Shan_2023, zhang2023pressure}, all with similar color sequences and different transition pressures. Similar Raman spectra to that in Ref.~\cite{Dias} have been observed in other experiments and have been attributed both to  LuH$_2$~\cite{luh2_raman, Color_change_LuH2} and  LuH$_3$~\cite{LuH3_ramman}, where both octahedral and tetrahedral interstitial sites are occupied by hydrogen. The difficulty to correctly characterize these materials is expected since both LuH$_2$ and LuH$_3$ should have similar lattice constants and consequently XRD patterns. 
	
	In this study, we give an explanation for these experimental results by combining density functional theory (DFT)~\cite{PBE,DFT+U,PseudoDojo,QE1,QE2,QE3,DFPT,DFPTU1,DFPTU2} with the stochastic self-consistent harmonic approximation (SSCHA)~\cite{SSCHA1,SSCHA2,SSCHA3,SSCHA4,SSCHA5} to study structural, vibrational, optical, and superconducting properties of both LuH$_2$ and LuH$_3$. First, we report the structural and elastic properties of lutetium hydrides, which we find  in reasonable agreement with experiments. The cubic LuH$_3$ shows imaginary phonon modes at 0 GPa and 0 K, even considering quantum anharmonic effects within the SSCHA. However, increasing pressure to 6 GPa and temperature to 300 K leads to a dynamically stable structure. We find that the 250 cm$^{-1}$ phonon peak observed in Raman could be a feature of the occupation of octahedral sites in LuH$_3$, only if quantum anharmonic effects are included. We have also calculated the optical reflectivity of both LuH$_2$ and LuH$_3$, and we only find the color change in LuH$_2$. LuH$_2$ has an undamped plasmon in the near-infrared region which gets blue-shifted towards the visible spectrum with increasing pressure, leading to a color change. Considering the differences between experiments and calculations, our work suggests that the parent structure of the material in Ref.~\cite{Dias}, as well as the materials observed in other experiments, is LuH$_2$ with slight doping of extra hydrogens in octahedral interstitial sites.

    \emph{Structural properties and phonon dispersion.}
	LuH$_2$ crystallizes in the cubic $Fm\bar{3}m$ structure, where Lu atoms form an fcc lattice and hydrogen atoms occupy tetrahedral sites. Experiments report a lattice constant of 5.033 \AA~\cite{LuH2_te_old} at 300 K or 5.028 \AA~\cite{LuH2_te_new} at 100 K. We have relaxed the structure of LuH$_2$ in static DFT, neglecting the zero-point energy, and within the SSCHA to account for it as well as anharmonicity. At 0 K we obtain the value of 5.07 \AA~within the SSCHA, which is a slight overestimation, expected  for the GGA exchange-correlation functional used. At ambient pressure LuH$_3$ crystallizes in the $P\bar{3}c1$ phase~\cite{LuH3_ramman,LuH3_structure}. With increasing pressure, it transforms to a cubic $Fm\bar{3}m$ structure, which is identical to LuH$_2$ with an additional hydrogen at the octahedral site. Experimentally, the phase transition between $P\bar{3}c1$ and $Fm\bar{3}m$ is reported at 12 GPa~\cite{LuH3_phase_transition} or 2 GPa~\cite{LuH3_ramman}. Our static DFT simulations predict the phase transition at 25 GPa (see Supplementary Material~\cite{supp_mat}). The disagreement between experiment and theory could be reconciled by accounting for anharmonic and zero-point motion effects within SSCHA. Due to the high computational cost for the less-symmetric $P\bar{3}c1$ phase, we have not performed these calculations. An additional reason for the discrepancy between calculations and experiments, as well as between different experiments, could be the existence of hydrogen vacancies in LuH$_3$ samples. 
	
	We have calculated the bulk modulus for  cubic LuH$_2$ and LuH$_3$ by fitting the static DFT energy to a third-order polynomial. Our results for LuH$_2$ give a bulk modulus of $B = 90.6$ GPa, and bulk modulus pressure derivative of $K_0 =  4.6$, which agree fairly well with the values found in Ref.~\cite{Dias} ($B = 88.6$ GPa and $K_0 = 4$). The reported experimental value for LuH$_3$,  $B = 89$ GPa~\cite{LuH3_bulkmodulus}, is very close to the value reported in Ref.~\cite{Dias}. The DFT results for cubic LuH$_3$ give a decent agreement for bulk modulus pressure derivative $K_0= 4$, but overestimate the value for bulk modulus $B = 104.7$ GPa. However, including anharmonic and quantum effects is going to lower the value of the bulk modulus, leading to a better agreement with the experiment for LuH$_3$. The literature also reports thermal expansion coefficients for LuH$_2$, 1.1985$\times 10^{-5}$ 1/K~\cite{LuH2_te_new} and 3.55$\times 10^{-5}$ 1/K~\cite{LuH2_te_old}. Within the SSCHA the volume thermal expansion coefficient is 2.77$\times 10^{-5}$ 1/K,  in good agreement with experiments.
	
\begin{figure}
		\centering
		\includegraphics[width=0.9\linewidth]{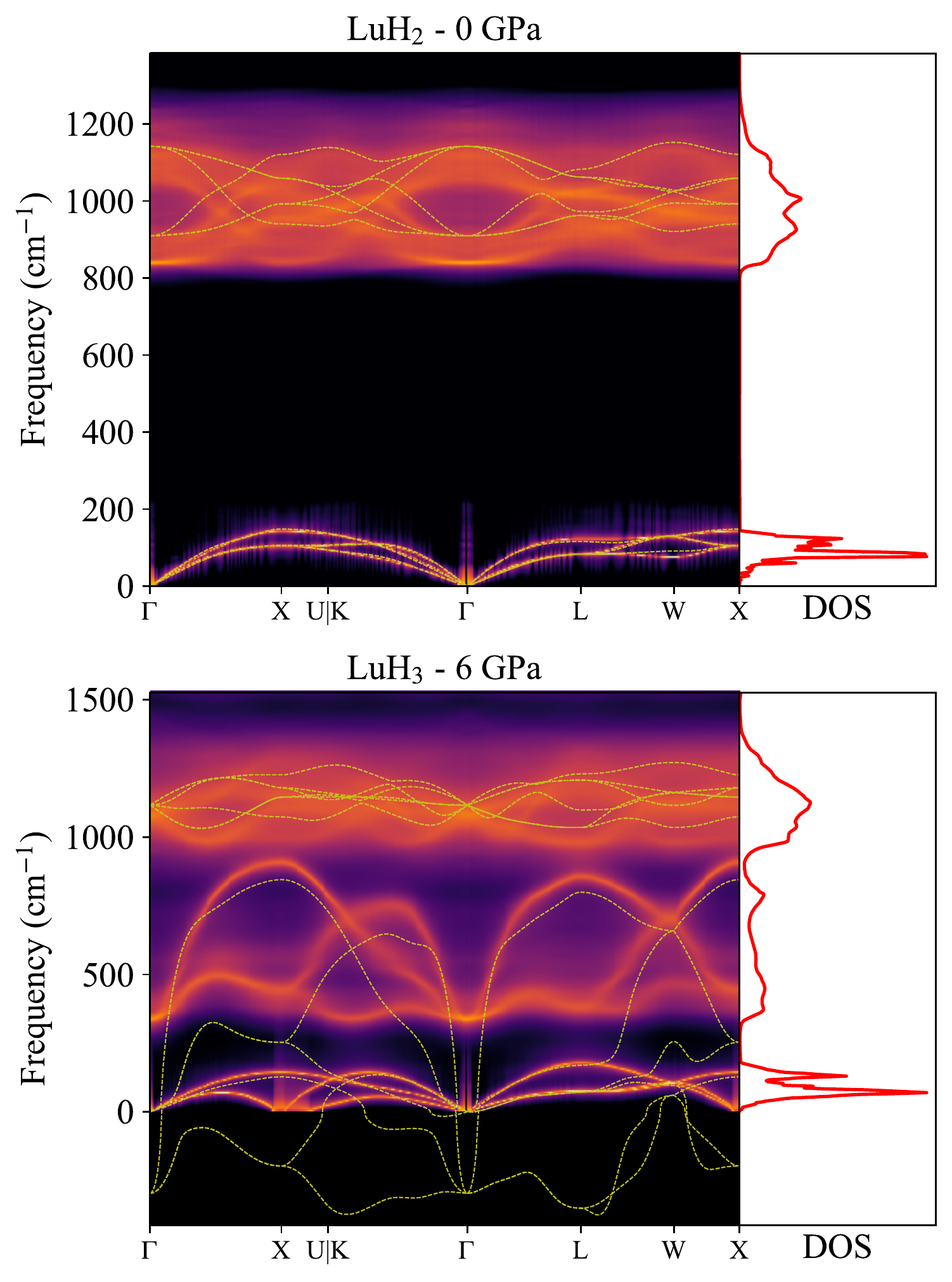}
		\caption{Phonon spectral function of LuH$_2$ (0 GPa) and LuH$_3$ (6 GPa) at 300 K. Side figures are showing phonon density of states obtained as a sum of the phonon spectral functions. The dashed lines are phonon frequencies calculated from DFPT (harmonic) force constants.}
		\label{phonons}
	\end{figure}

	We calculate the phonon properties of  lutetium hydrides at the harmonic level within density functional perturbation theory (DFPT)~\cite{DFPT, DFPTU1, DFPTU2} and including anharmonicity within the SSCHA (see Fig. \ref{phonons}). LuH$_2$ is very harmonic and both harmonic DFPT and SSCHA calculations give similar phonon band structures. Including dynamical effects through the dynamical bubble approximation~\cite{TDSSCHA} leads to moderate phonon spectral broadening and small temperature lineshifts. In LuH$_2$ phonon modes are separated into two groups, low-frequency modes (below 200 cm$^{-1}$), which have mostly lutetium character, and high-frequency (around 1000 cm$^{-1}$) optical modes associated to hydrogen atoms in the tetrahedral sites. These bands are separated with a large frequency gap and cannot explain the Raman spectra in lutetium hydrides, which show phonon modes inside this gap~\cite{Dias, luh2_raman, LuH3_ramman}.
	
	Cubic LuH$_3$ is dynamically unstable at 0 GPa~\cite{LuH3_SSCHA}, showing imaginary phonon frequencies throughout the Brillouin zone. If we apply pressure of 6 GPa and increase the temperature to 300 K the structure stabilizes due to anharmonicity, as evidenced by the SSCHA free energy Hessian dispersion (see Supp.~Mat.~\cite{supp_mat}). This compound is therefore strongly anharmonic, as it happens for other fcc metals with H atoms in octahedral sites~\cite{SSCHA2,SSCHA3,Meninno2023AbInitio}. The spectral function calculation reveals there might be an instability around the X point of the Brillouin zone. A closer inspection of the spectral function at this point reveals strong softening. However, the spectral weight goes to zero as $\omega$ goes to zero and thus it is not an instability. Phonons of this phase strongly resemble those of LuH$_2$ with additional phonon branches inside the acoustic-optical frequency gap. These phonons are associated with hydrogen atoms in octahedral sites and are not Raman active. However, if there is enough disorder in the experimental sample, they could be the source of the 250 cm$^{-1}$ peak. In our calculations, octahedral site modes are very sensitive to the applied pressure, considerably more than in the experiment~\cite{LuH_color_change_higher_pressure}. They stiffen above 300 cm$^{-1}$ at 6 GPa and thus cannot be conclusively identified as the origin of the 250 cm$^{-1}$ peak. On the other hand, the octahedral modes are not at all present in LuH$_2$ and thus are a strong indication that we have at least partial occupation of octahedral sites in the samples. Alternatively, the Raman signal at 250 cm$^{-1}$ from Ref.~\cite{Dias} could be explained by nitrogen-dominated modes, which should exist in this frequency range. However, this does not explain why these modes are also observed in undoped LuH$_2$ and LuH$_3$ samples~\cite{luh2_raman,Color_change_LuH2,LuH3_ramman}.
	
	\begin{figure}[h]
		\centering
		\includegraphics[width=0.9\linewidth]{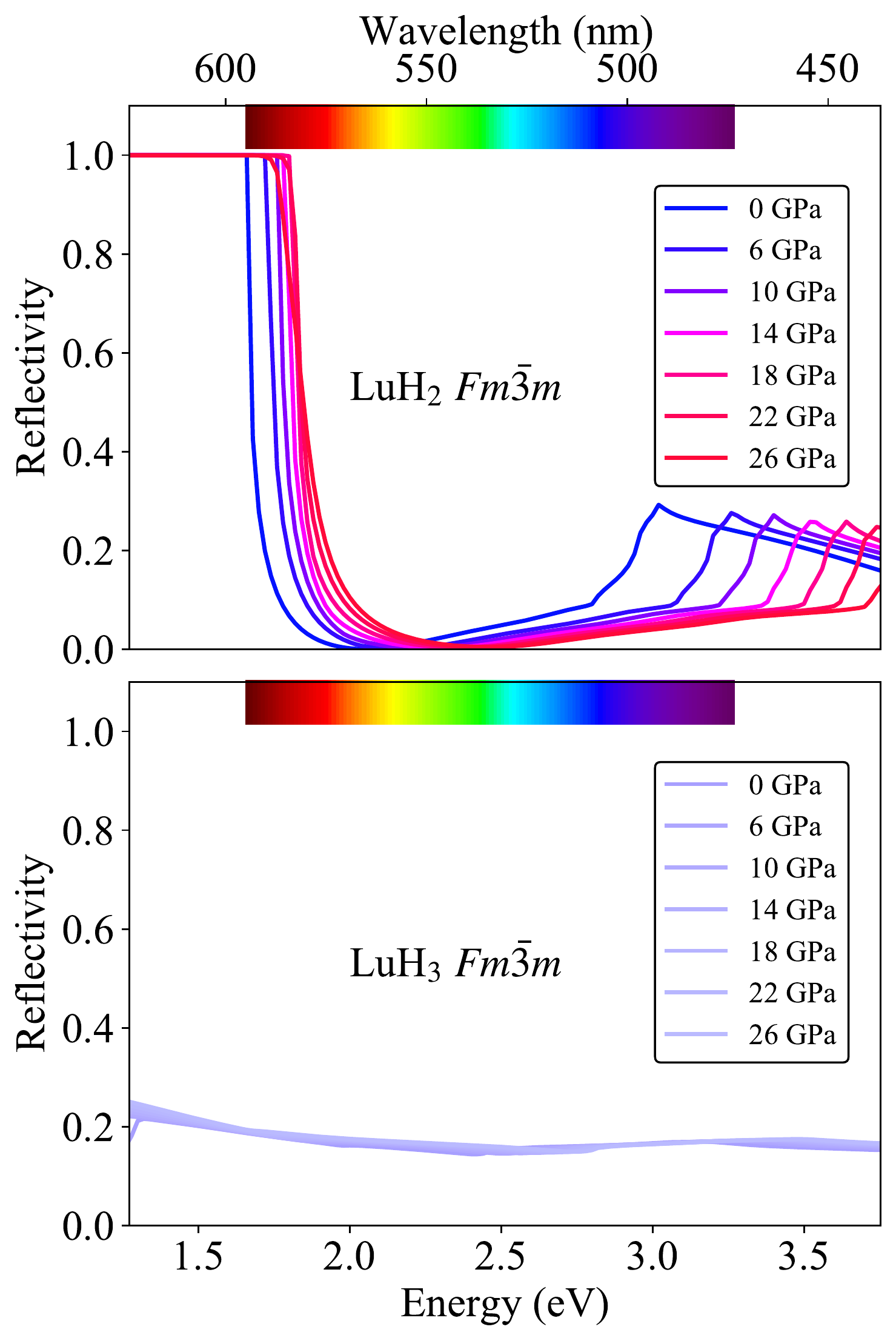}
		\caption{Calculated reflectivities of LuH$_2$ (top panel) and LuH$_3$ (bottom panel). The line color represents the perceived color calculated from the reflectivity data.}
		\label{reflectivities}
	\end{figure}

    \emph{Optical properties.}
	The color change of the sample in the experiment in Ref.~\cite{Dias} is very intriguing. It has been reproduced multiple times in different experiments but with different transition pressures for the color changes~\cite{Color_change_LuH2,LuH_color_change_higher_pressure}. To model this behavior of lutetium hydrides, we performed electronic band structure calculations and, afterwards, calculated the optical dielectric function within the random-phase approximation by means of Wannier interpolation. From the dielectric function, it is straightforward to obtain the reflectivity of the sample using the Fresnel equation. The actual color of the sample is calculated by converting the reflectivity using color-matching functions to a standard RGB format~\cite{Prandini2019}. The results are presented in Fig.~\ref{reflectivities}. The color of LuH$_2$ is changing from blue to red in the span of 26 GPa, which is in agreement with one of the experiments~\cite{LuH_color_change_higher_pressure}. Since in the experiment the sample is loaded in the diamond anvil cell, in the Fresnel equations we assume that the light reflects from the diamond. This is probably the source of disagreement between our calculations with some of the recent ones~\cite{monserrat}, where the medium is assumed to be vacuum, which resulted in color transitions at much higher pressures. On the other hand, the cubic LuH$_3$ does not show any color change in the considered pressure range. 
	
\begin{figure*}[t]
		\centering
		\includegraphics[width=0.99\linewidth]{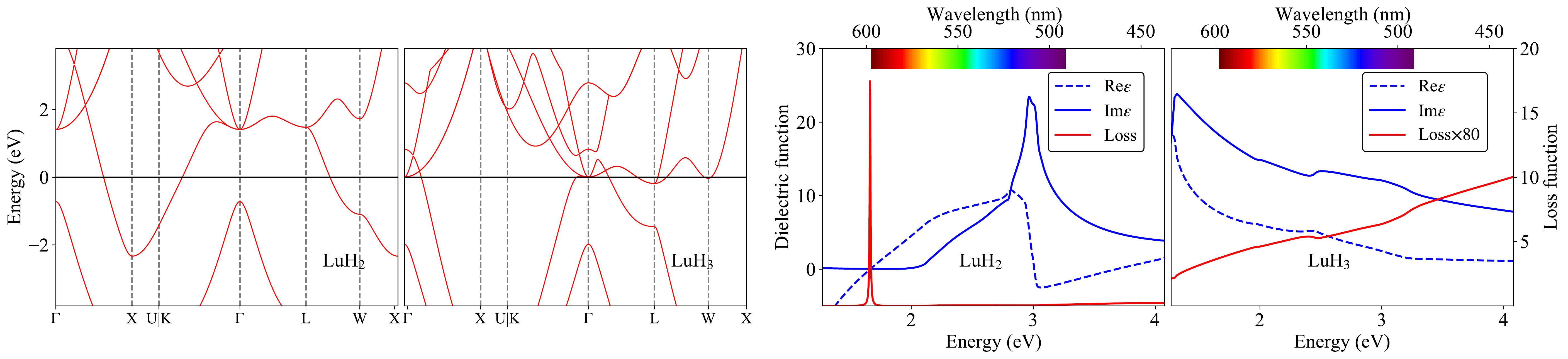}
		\caption{Electronic band structure of LuH$_2$ and LuH$_3$ at 0 GPa. The real and imaginary parts of the dielectric function $\varepsilon(\omega)$ calculated in the random phase approximation are given as dashed and full blue lines, respectively. The energy loss function, $-\mathrm{Im}[\varepsilon(\omega)^{-1}]$, is given as full red line.}
		\label{plasmon}
	\end{figure*}

	The reason for the color change in LuH$_2$ is the existence of an undamped interband plasmon~\cite{PhysRevB.81.205105} in the near-infrared region and a lack of interband electronic optical transitions below 2 eV (see Supp.~Mat.~\cite{supp_mat}). The onset of interband transitions makes the imaginary part of the dielectric function soar from zero above approximately 2 eV, making the real part cross the zero value at lower energies due to Kramers-Kronig relations. Thus, both  real and imaginary parts of the dielectric function are zero at the same energy and the energy-loss function has a Dirac delta peak at this energy, resulting in an undamped plasmon peak (see Fig.~\ref{plasmon}). The presence of the plasmon is responsible for suppressing the large reflectivity in the far infrared, making consequently the sample blue. As pressure is increased, this plasmon blue shifts, and the highly reflecting region enters the visible range, making the overall color of the sample red and shiny. 
	Interband transitions are present at all energies in LuH$_3$ and hence it does not have any interband plasmon in the optical range. Its optical properties barely change with pressure and there is no color change. Additionally, LuH$_3$ samples should reflect much less light than LuH$_2$.
	
	In the study we are not including the effects of electron-phonon coupling on optical properties of LuH$_2$ and LuH$_3$. In LuH$_2$, electron-phonon interaction will lead to the broadening of the electronic band structure in Fig.~\ref{plasmon} and push the onset of available interband transition to slightly lower energies. This will not lead to a large change in reflectivity and the color change should happen at similar pressures. In LuH$_3$ the effect of electron-phonon coupling should be even smaller due to the fact that electronic interband transitions are already allowed. 

    \emph{Discussion}
	At this moment, we have contradictory findings. While the Raman signal can only be explained by LuH$_3$, the color changes induced by pressure only exist in LuH$_2$. As we mentioned before,  XRD cannot distinguish between these two materials, while the calculated lattice constant is overestimated in DFT/SSCHA for both of the models. The calculated elastic constants are similar enough that we cannot say with certainty which material appears in the experiments. Some previous experiments suggested that the stoichiometry of lutetium hydride is between lutetium dihydride and lutetium trihydride~\cite{Daou_1988}. Indeed, at 1 GPa doping the lutetium dihydride with H (Lu$_4$H$_9$) puts it only 12 meV/atom above the convex hull, which is significantly lower than LuH$_3$ (82 meV/atom). The crystal structure search that we have performed (see Supplementary Material) with Lu$_4$H$_9$ stoichiometry suggests that extra H should go into the octahedral sites. This configuration will give rise to the octahedral site optical phonon modes that are the best candidate for the source of the Raman peak at 250 cm$^{-1}$ in experiments. However, this addition also pushes the plasmon deeper into the infrared region, which means that the color change would be induced at a higher pressures than in pure LuH$_2$. 
	
	We have also considered the case of LuH$_2$ with vacancies on the tetrahedral site. Our crystal structure predictions reveal that in this case (Lu4H7), the energy above the convex hull is slightly higher compared to hydrogen-doped LuH2 (Lu4H9), specifically by 27 meV as opposed to 12 meV. The Lu$_2$H$_3$ structure is red already at 0 GPa and with increasing pressure changes color to yellow/orange. Additionally, it is not able to explain the 250 cm$^{-1}$ Raman peak since it does not have phonon modes in the acoustic-optical phonon gap. 
	
	Finally, we have also calculated the superconducting properties of lutetium dihydride and trihydride using isotropic Migdal-Eliashberg equations. In agreement with experiments~\cite{Daou_1984,Daou_1988} and recent theoretical work~\cite{LuH3_SSCHA,monserrat} we do not find superconductivity in LuH$_2$. Using the stable structure of LuH$_3$ at 6 GPa and 300 K, we find that LuH$_3$ should have a critical temperature of 19 K, which is well below the temperature needed to stabilize it. We find that the anharmonicity of phonon modes plays a large role in determining the critical temperature, as it changes by a 50\% depending on the method used to calculate the Eliashberg spectral function $\alpha^2F(\omega)$. The modes that contribute significantly to $\alpha^2F(\omega)$ are the octahedral modes, as it happens in palladium hydrides~\cite{Meninno2023AbInitio}, which explains the existence of superconductivity in LuH$_3$ and not in LuH$_2$.
	
	In conclusion, we have performed a first-principles study of the physical properties of lutetium hydrides in order to recognize the parent structure of the material synthesized in Ref.~\cite{Dias}. We find that both LuH$_2$ and LuH$_3$ have similar structural and elastic properties and are thus indistinguishable in XRD experiments. Their phonon band structures, however, are considerably different, with only LuH$_3$ being able to explain experimental findings. LuH$_3$ is dynamically unstable at 0 K and 0GPa but stabilizes at pressures above 6 GPa above room temperature thanks to anharmonic effects. The color change observed in many experiments is only a feature of LuH$_2$ and does not take place in LuH$_3$. For these reasons, we believe that the structure which is synthesized in most experiments is LuH$_2$ with extra hydrogens in octahedral sites. Isotropic Migdal-Eliashberg calculations show that LuH$_2$ is not a superconductor, while LuH$_3$ has a modest critical temperature, significantly less than what is reported in Ref.~\cite{Dias}.

	\bigskip
 
	This work is supported by the European Research Council (ERC) under the European Unions Horizon 2020 research and innovation program (grant agreement No. 802533 and grant agreement No. 946629) and the Department of Education, Universities and Research of the Eusko Jaurlaritza and the University of the Basque Country UPV/EHU (Grant No. IT1527-22). We acknowledge PRACE for awarding us access to Lumi located in CSC’s data center in Kajaani, Finland.
	
	\bibliography{arxiv_version}
	\onecolumngrid
	\renewcommand{\figurename}{Supplementary Figure}
	\renewcommand{\tablename}{Supplementary Table}
	\setcounter{figure}{0}
	\newpage
	\newpage
	\newpage

\section*{Supplementary Material: \\ {\it Ab initio} study of the structural, vibrational and optical properties of potential parent structures of nitrogen-doped lutetium hydride}
\subsection{Computational details}

Density functional theory (DFT)~\cite{DFT+U} and density functional perturbation theory (DFPT)~\cite{DFPT,DFPTU1,DFPTU2} calculations with the Perdew-Burke–Ernzerhof parametrization for the generalized gradient approximation~\cite{PBE} were performed using Quantum Espresso software package~\cite{QE1,QE2,QE3}. Ions were represented using PseudoDojo~\cite{PseudoDojo} generated norm-conserving pseudopotentials. Electronic-wave functions were defined in the plane-wave basis with energy cutoff of 100 Ry, while the energy cutoff for the charge density was 400 Ry. Hubbard correction (DFT+$U$)~\cite{DFT+U} was used for Lu \textit{f} electrons with $U = 5.5$ eV. The $\mathbf{k}$ point grid used to sample the electronic states were 14$\times 14\times 14$ for both LuH$_2$ and LuH$_3$. Due to the metallic nature of $Fm\bar{3}m$ LuH$_2$ and LuH$_3$, we used Gaussian smearing for electronic states of 0.015 Ry for self-consistent calculations.

Stochastic self-consistent harmonic approximation (SSCHA)~\cite{SSCHA1,SSCHA2,SSCHA3,SSCHA4,SSCHA5} calculations were done on $3\times 3\times 3$ supercells. We used 50 configurations of random atomic positions according to SSCHA auxiliary force constants per populations for LuH$_2$ and 200 configurations for LuH$_3$. The third-order force constants and hessian of free energy were calculated using 1000 configurations in both cases. 

The calculation of superconducting properties of LuH$_3$ was done for the stable structure at 6 GPa and 300 K. We used $8\times 8\times 8$ $\mathbf{q}$ point grid to sample vibrational properties and $42\times 42\times 42$ $\mathbf{k}$ point grid to sample electronic states. Double delta averaging of electron-phonon matrix elements over the Fermi surface was done with 0.008 Ry Gaussian smearing. Critical temperature was calculated using isotropic Migdal-Eliashberg equations using SSCHA auxiliary phonons, SSCHA hessian phonons and full anharmonic phonon spectral functions~\cite{our_hydrogen}. The cutoff for Matsubara frequencies was 0.117 Ry (10 times the highest phonon frequency), and the value of the reduced Coloumb repulsion was $\mu^* = 0.1$.

In order to calculate the optical properties of lutetium hydrides, we first constructed maximally localized Wannier functions (MLWFs) using the \textsc{Wannier90} code package~\cite{Pizzi_2020}.
In the case of LuH$_{2}$ (LuH$_{3}$) at all pressures, starting from a set of 20 (22) spin-degenerate bands, we constructed 18 (19) disentangled MLWFs spanning the nine high-energy fully occupied bands and the nine (ten) low-energy partially occupied and unoccupied bands using one $s$ trial orbital centered on H atoms, as well as one $s$, one $p$, one $d$ and one $f$ on the Lu atom.
In the case of Lu$_{2}$H$_{3}$ (Lu$_{2}$H$_{5}$) at all pressures, starting from a set of 40 spin-degenerate bands, we constructed 35 (37) disentangled MLWFs spanning the eighteen (nineteen) high-energy fully occupied bands and the seventeen (eighteen) low-energy partially occupied and unoccupied bands.
For Lu$_{2}$H$_{3}$, we used one $p$ trial orbital centered on H atoms, as well as one $s$, one $d$ and one $f$ on Lu atoms.
For Lu$_{2}$H$_{5}$ we used one $s$ trial orbital centered on H atoms, as well as one $s$, one $p$, one $d$ and one $f$ on Lu atoms.

Once our Wannier basis is converged, we then computed the linear optical response within the independent-particle approximation (IPA) by means of Eqs.~(B1a)-(B1b) of Ref.~\cite{PhysRevB.107.205101} using Wannier interpolation.
To that end, we used the schemes described in Ref.~\cite{PhysRevB.74.195118} for the calculation of interband dipole matrix elements and Berry curvatures, and the scheme described in Ref.~\cite{PhysRevB.102.205123} for the calculation of the effective mass inverse tensor, together with the scheme described in Ref.~\cite{PhysRevB.75.195121} for the calculation of the velocity matrix elements.
To obtain well-converged optical spectra, we used dense $\mathbf{k}$ point interpolated meshes of $200\times200\times200$ for all structures at all pressures.
With respect to the linear Kohn-Sham (KS) interband polarizability tensor expression in Eq.~(B1a) of Ref.~\cite{PhysRevB.107.205101}, we employed an adaptative scheme~\cite{PhysRevB.75.195121} for setting the broadening parameter $\eta$ in the energy denominator $1/(\omega_{mn}-\tilde{\omega})$ with the complex frequency $\tilde{\omega}=\omega+i\eta/\hbar$.
Regarding the linear KS intraband conductivity tensor expression in Eq.~(B2a) of Ref.~\cite{PhysRevB.107.205101}, we set $\gamma=0.01~\mathrm{eV}$ in the energy denominator of the Drude-peak term $1/\tilde{\omega}$ with the complex frequency $\tilde{\omega}=\omega+i\gamma/\hbar$.
The occupation factors  were evaluated at room temperature $(T=300~\mathrm{K})$ for all structures at all pressures.

Finally, from the linear KS optical response, we calculated the optical dielectric function within the random-phase approximation without crystal-local fields, by setting $\alpha_{\mathrm{LRC}}=0$ through Eq.~(33) of Ref.~\cite{PhysRevB.107.205101} and averaging over Cartesian coordinates $a$: $\varepsilon_{\mathrm{RPA}}(\omega)=\sum_{a}\varepsilon^{aa}_{\mathrm{RPA}}(\omega)/3$. The resulting reflectivity $R$ is then calculated using the Fresnel equation: $R(\omega) = \lvert\frac{\sqrt{\varepsilon_{\mathrm{RPA}}(\omega)} - n_d}{\sqrt{\varepsilon_{\mathrm{RPA}}(\omega)} + n_d}\rvert^2$, where $n_d$ is the refraction index of diamond anvil cell, which we assume to be 2.33.
The electron energy loss (EEL) function is calculated as $-\mathrm{Im}[\varepsilon_{\mathrm{RPA}}^{-1}(\omega)]$.

To construct the convex hull of Lu-H, we consider the  $Fm\bar{3}m$ LuH$_2$,   $Fm\bar{3}m$ LuH$_3$, and $P\bar{3}c1$  LuH$_3$ screened from the Materials Project database. The former two have been extensively studied in our main text. In addition, the Lu$_2$H$_5$ which denotes hydrogen doped at the octahedral site of LuH$_2$ has been included. The other two binary hydrides, Lu$_4$H$_7$ and Lu$_4$H$_9$, are obtained from high-throughput crystal structure predictions with the CALYPSO code~\cite{Wang2010,Wang2012}. The crystal structure search was performed over 1500 crystal structures up to 1 formula unit of Lu$_4$H$_7$ and Lu$_4$H$_9$. We used the Vienna Ab initio Simulation Package
(VASP)~\cite{Kresse-PRB-1996,Kresse1996} as the DFT calculator. 
The generalized gradient approximation in the form of the Perdew, Burke, and Ernzerhof exchange-correlation functional~\cite{PBE} has been used with a Hubbard $U$ correction ($U$ = 5.5 eV) in the Dudarev’s form~\cite{DFT+U}. The plane wave energy cutoff was set to 450 eV during crystal structure prediction with the energy criteria of 1$\times$10$^{-5}$ eV and the force criteria of 0.001 eV/\AA.
The \textbf{k}-grid is generated based on the specific structure by Pymatgen~\cite{Jain2011_pymatgen} with a relatively high grid density of 60 per \AA$^{-3}$ of reciprocal cell. All the calculations relevant to the convex hull are evaluated at 1 GPa.

\newpage
\subsection{SSCHA and harmonic phonon band structure}

In Supplementary Fig.~\ref{phonons1} we show band dispersion of harmonic phonon modes and free energy hessians for LuH$_2$ and LuH$_3$ calculated from DFPT for SSCHA structures at (0 GPa, 0K) and (6 GPa, 300 K) respectively. LuH$_2$ has very small differences between two methods and mostly only for optical phonon modes. On the other hand, we can see that DFPT predicts large number of imaginary phonon modes for LuH$_3$. SSCHA shows that once we include temperature and quantum effects these instabilities get cured and the structure is stable. There is a very low-laying acoustic phonon mode in the vicinity of reciprocal points U (0.625, 0.250, 0.625) and K (0.375, 0.375, 0.750). 

\begin{figure}[h]
	\centering
	\includegraphics[width=0.9\textwidth]{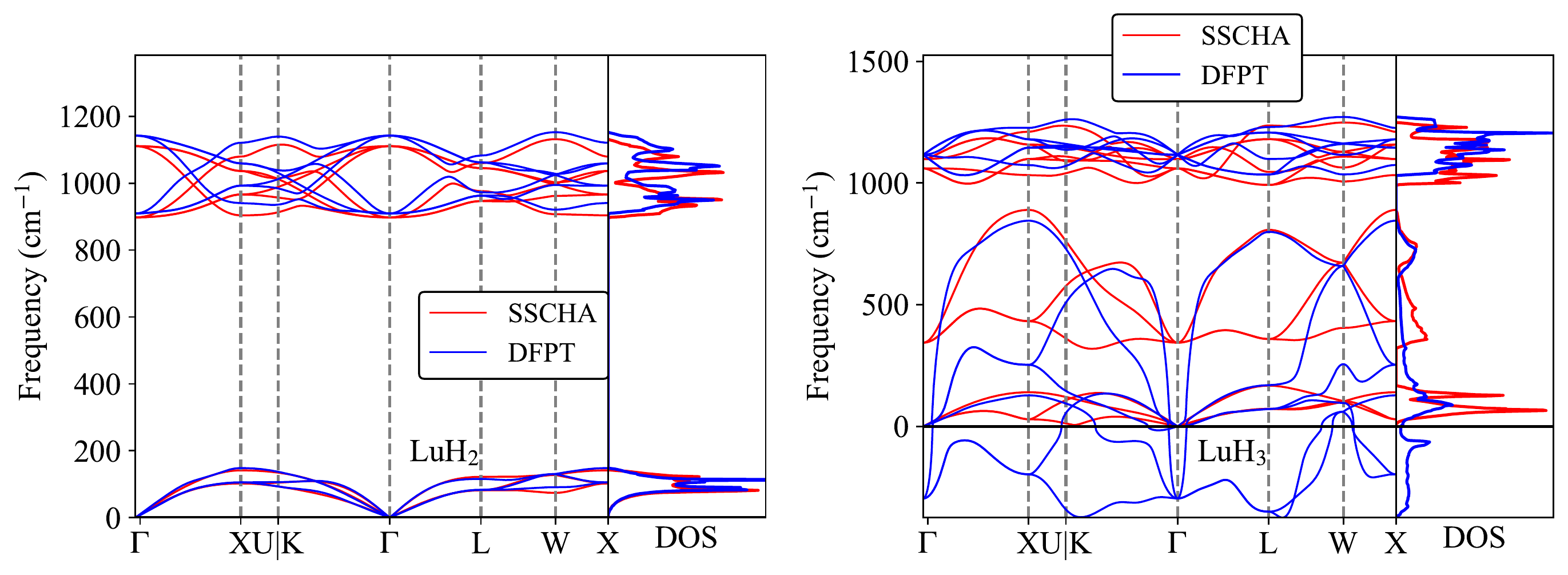}
	\caption{Band structure along the high symmetry lines of harmonic phonon modes and free energy hessians for LuH$_2$ and LuH$_3$ calculated from DFPT for SSCHA structures at (0 GPa, 0K) and (6 GPa, 300 K) respectively.}
	\label{phonons1}
\end{figure}

\newpage
\subsection{Structural and elastic properties of lutetium hydrides}

\begin{figure}[h]
	\centering
	\includegraphics[width=0.9\textwidth]{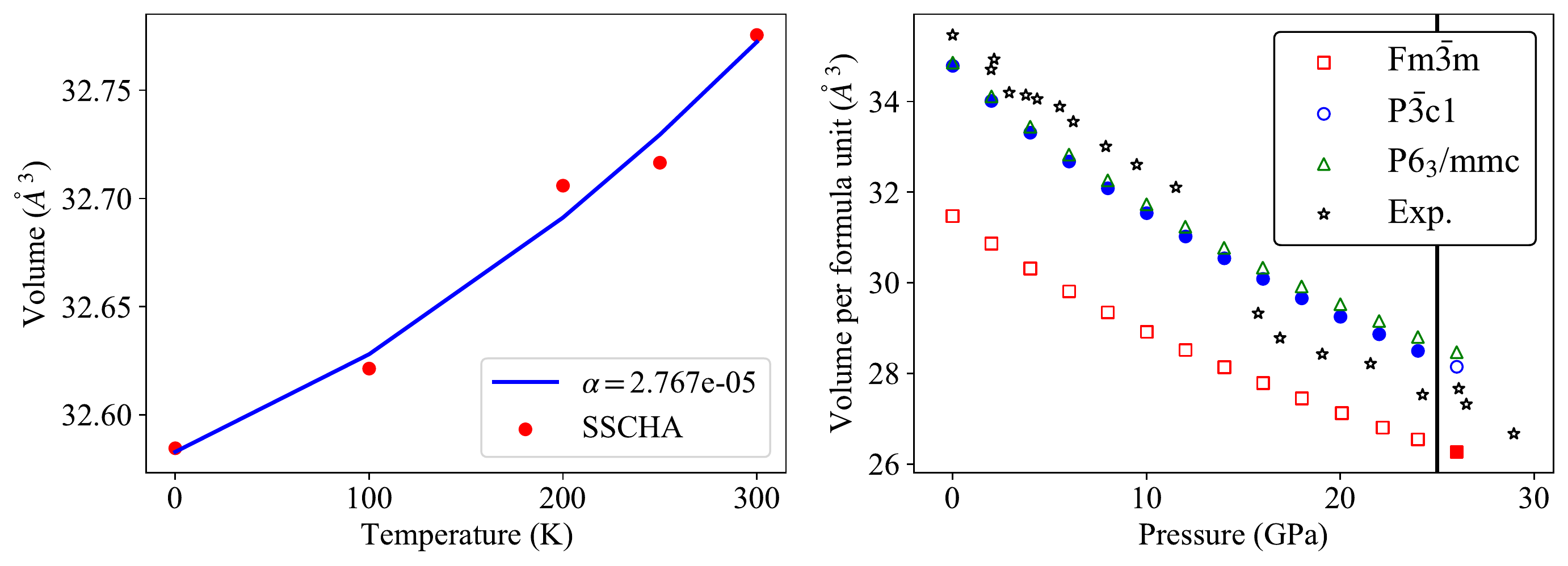}
	\caption{Left panel: Thermal expansion of LuH$_2$ calculated in SSCHA. Right panel: The phase diagram of LuH$_3$ calculated in DFT. The full symbols represent structures with lowest enthalpy.}
	\label{te}
\end{figure}

In left panel of Fig.~\ref{te} we show the volumes of LuH$_2$ calculated at different temperatures using SSCHA. The volumes are obtained by relaxing structure of LuH$_2$ inside SSCHA until the averaged pressure is below the statistical error of the calculated pressure. We then fit the set of $(V,T)$ points to the polynomial of order 2 ($V(T) = V_0 + aT + bT^2$). The thermal expansion coefficient at 300 K is then calculated as the analytic derivative of the obtained polynomial ($\alpha = \frac{1}{V_0}\frac{\partial V}{\partial T} = \frac{1}{V_0}(a + 2bT)$).

The right panel in Fig.~\ref{te} represent phase diagram of LuH$_3$ calculated in DFT. We consider three phases: $Fm\bar{3}m$, $P\bar{3}c1$, and hexagonal $P6_3/mmc$. The full symbols denote structures with the lowest enthalpy. This means that in DFT the expected transition pressure from $P\bar{3}c1$ to $Fm\bar{3}m$ in LuH$_3$ would be 25 GPa.

\begin{figure}[h]
	\centering
	\includegraphics[width=0.9\textwidth]{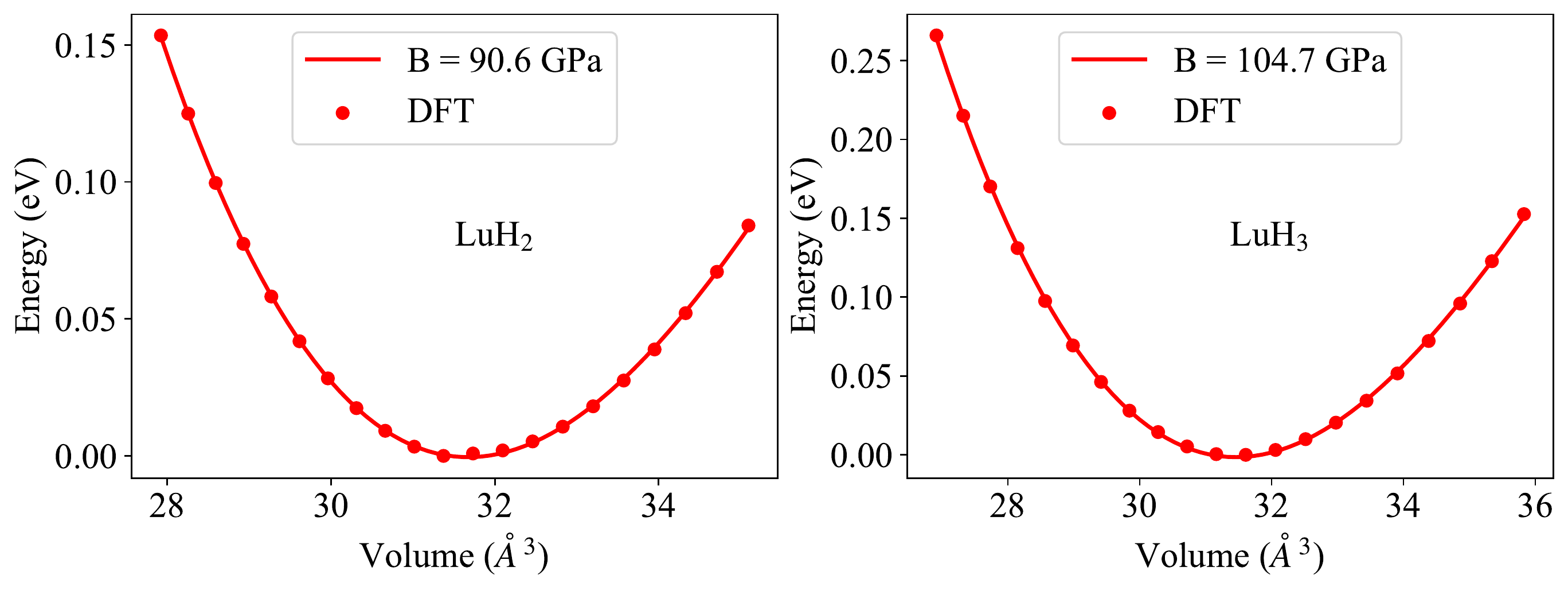}
	\caption{Equation of state for Fm$\bar{3}$m LuH$_2$ and LuH$_3$ in DFT.}
	\label{bulk_modulus}
\end{figure}

In Fig.~\ref{bulk_modulus} we show the equation of state (EOS) for LuH$_2$ and LuH$_3$. Calculated bulk modulus was obtained by fitting the said EOS to the third polynomial and then obtaining bulk modulus analytically.

\newpage
\subsection{Optical properties}
\begin{figure}[h]
	\centering
	\includegraphics[width=0.85\linewidth]{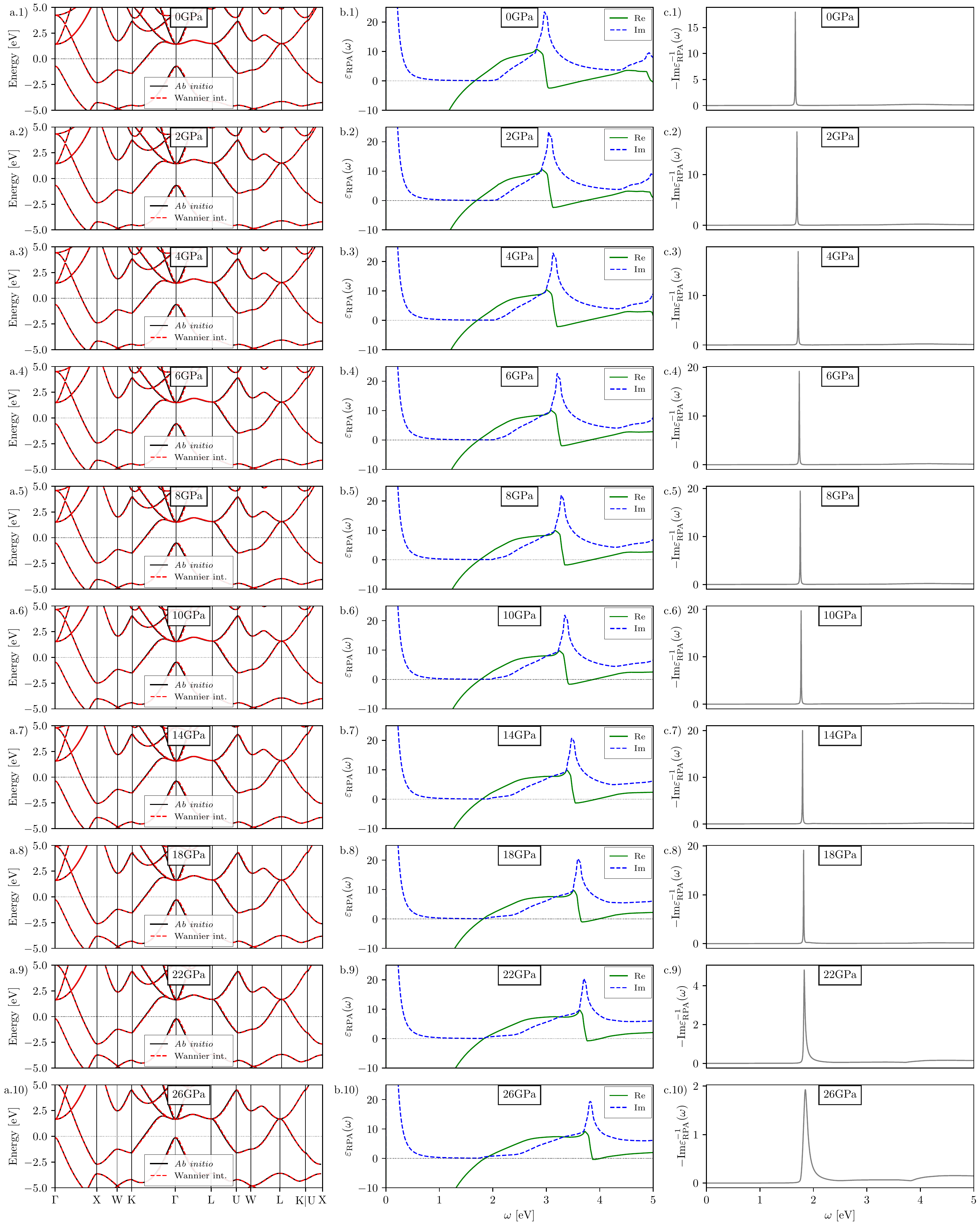}
	\caption{(a.1)-(a.10) DFT (solid black lines) and Wannier-interpolated (dashed red lines) energy bands, (b.1)-(b.10) real (solid green line) and imaginary (dashed blue line) parts of the dielectric function, and (c.1)-(c.10) EEL function of LuH$_{2}$ at 0, 2, 4, 6, 8, 10, 14, 18, 22 and 26 GPa and room temperature.}
	\label{luh2_bands}
\end{figure}

\begin{figure}
	\includegraphics[width=0.5\linewidth]{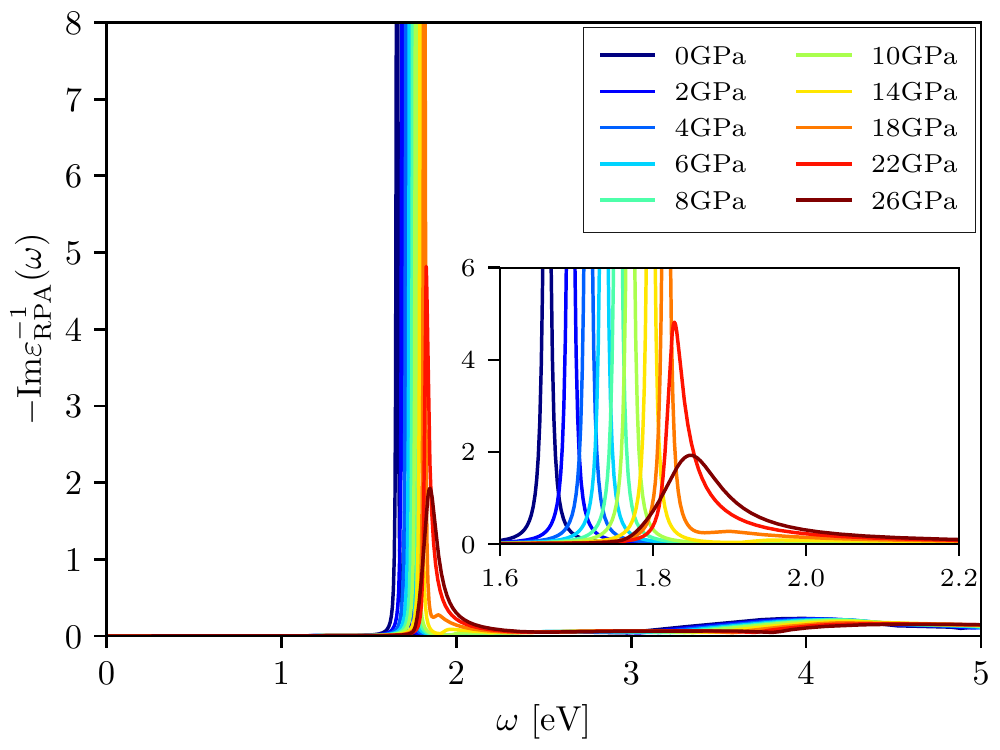}
	\caption{Evolution of the EEL spectrum as a function of pressure for LuH$_{2}$ at room temperature in the energy range $[0;5]~\mathrm{eV}$. The inset focuses on the evolution of the plasmon-peak position and shape in the energy range $[1.6;2.2]~\mathrm{eV}$, \textit{i.e.}~the border region between the near-infrared and the red color spectrum. The color code is a guide for the eye.}
	\label{plasmon1}
\end{figure}

\begin{figure}[h]
	\includegraphics[width=0.9\linewidth]{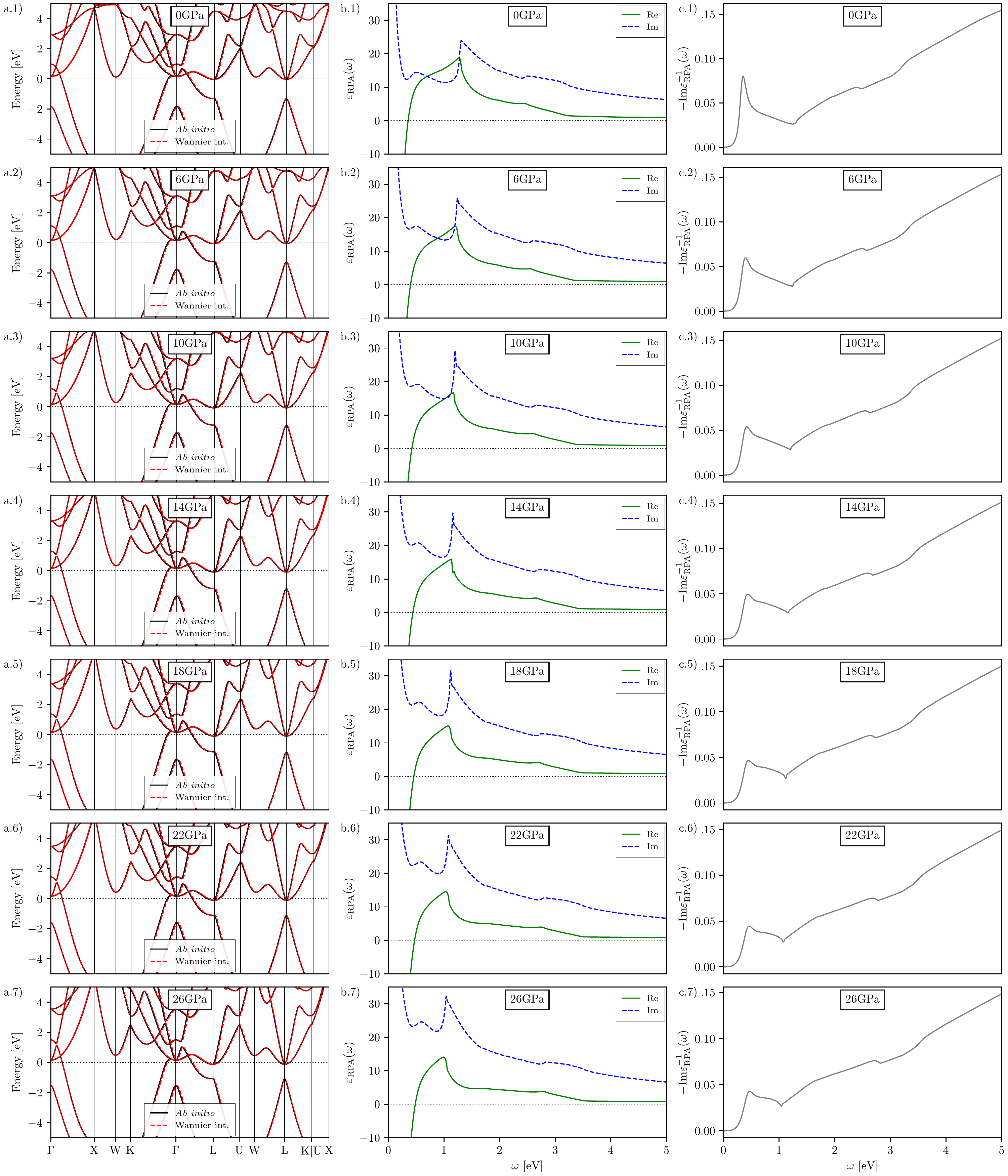}
	\caption{(a.1)-(a.7) DFT (solid black lines) and Wannier-interpolated (dashed red lines) energy bands, (b.1)-(b.7) real (solid green line) and imaginary (dashed blue line) parts of the dielectric function, and (c.1)-(c.7) EEL function of LuH$_{3}$ at 0, 6, 10, 14, 18, 22 and 26 GPa and room temperature.}
	\label{luh3_bands}
\end{figure}

\begin{figure}[h]
	\includegraphics[width=\linewidth]{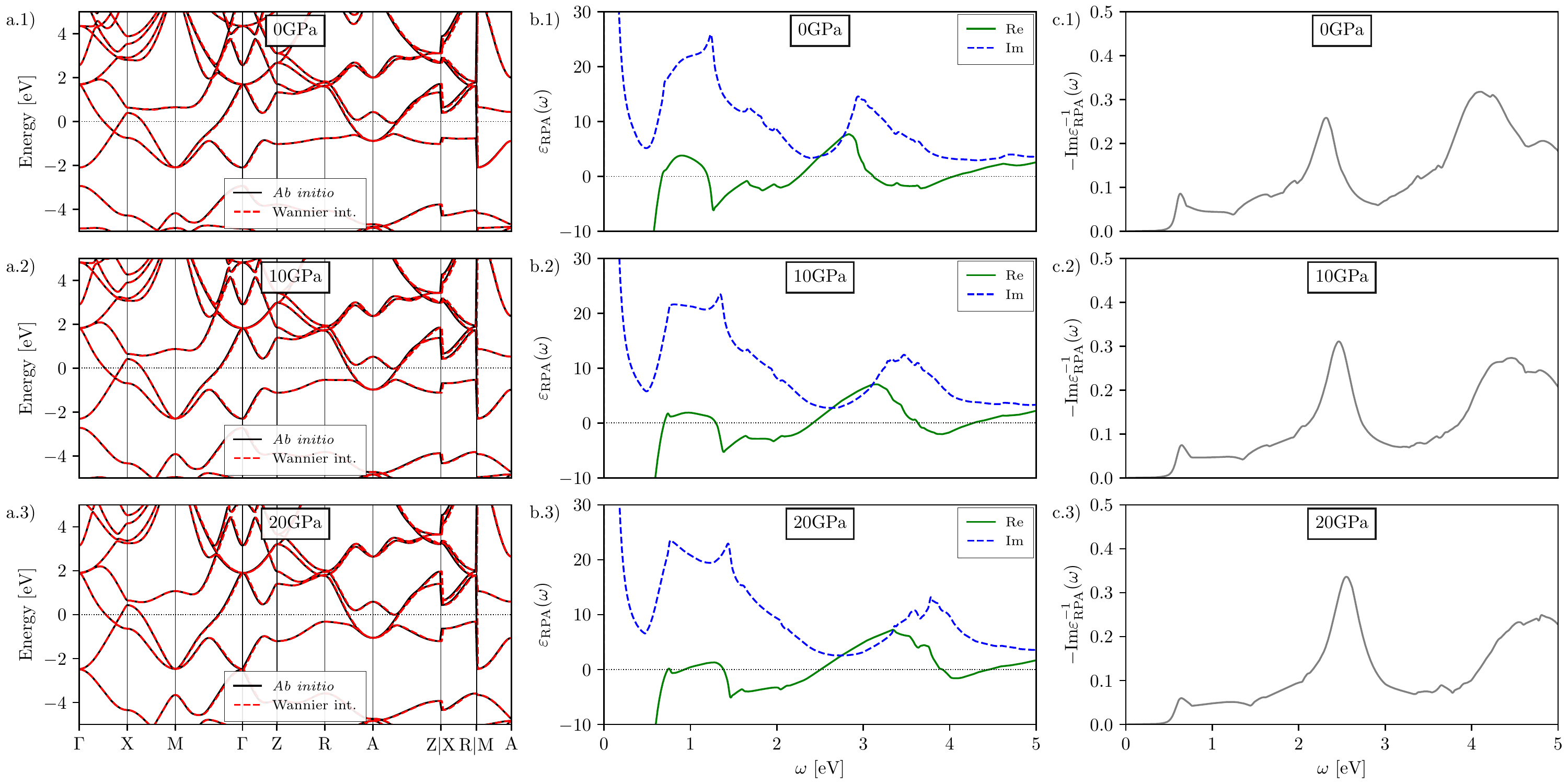}
	\caption{(a.1)-(a.7) DFT (solid black lines) and Wannier-interpolated (dashed red lines) energy bands, (b.1)-(b.7) real (solid green line) and imaginary (dashed blue line) parts of the dielectric function, and (c.1)-(c.7) EEL function of Lu$_{2}$H$_{3}$ at 0, 10 and 20 GPa and room temperature.}
	\label{lu2h3_bands}
\end{figure}

\begin{figure}[h]
	\includegraphics[width=\linewidth]{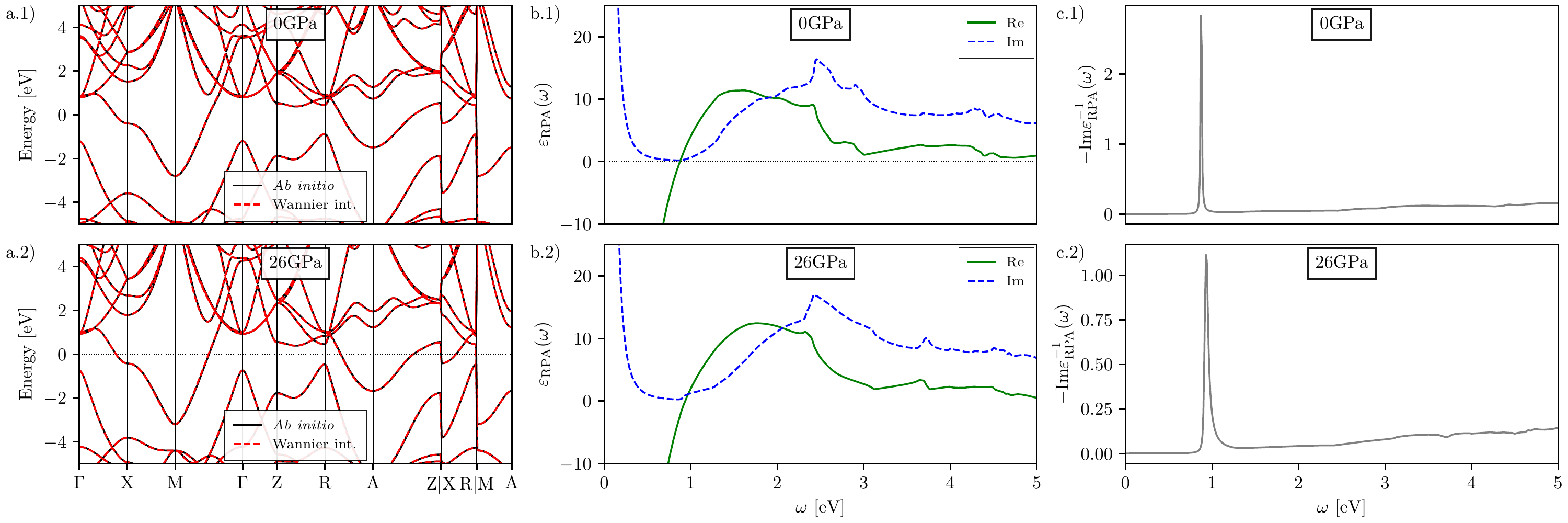}
	\caption{(a.1)-(a.2) DFT (solid black lines) and Wannier-interpolated (dashed red lines) energy bands, (b.1)-(b.2) real (solid green line) and imaginary (dashed blue line) parts of the dielectric function, and (c.1)-(c.2) EEL function of Lu$_{2}$H$_{5}$ at 0 and 26 GPa and room temperature.}
	\label{lu2h5_bands}
\end{figure}

In this section, we present the calculations of the electronic band structure, the optical dielectric function and the EEL function of LuH$_{2}$ (see Fig.~\ref{luh2_bands}), LuH$_{3}$ (see Fig.~\ref{luh3_bands}), Lu$_{2}$H$_{3}$ (see Fig.~\ref{lu2h3_bands}) and Lu$_{2}$H$_{5}$ (see Fig.~\ref{lu2h5_bands}) at different pressures and room temperature. Regarding the electronic structure, DFT and Wannier-interpolated bands are represented together in order to visually show the high-quality wannierizations used for calculating dielectric functions. In the case of LuH$_{2}$, in Fig.~\ref{plasmon1} we show the evolution of the undamped interband plasmon with pressure. With increasing pressure, the plasmon peak position is blue-shifted from the near-infrared region to the red color region. Indeed, this trend explains the change in the reflectivity spectrum, and ultimately the color change observed experimentally.
\clearpage
\subsection*{LuH$_{2}$}
\clearpage
\newpage
\subsection*{LuH$_{3}$}
\newpage
\subsection*{Lu$_{2}$H$_{3}$}
\newpage
\subsection*{Lu$_{2}$H$_{5}$}

\newpage
\subsection{Superconductivity in LuH$_3$}

\begin{figure}[h]
	\centering
	\includegraphics[width=0.9\textwidth]{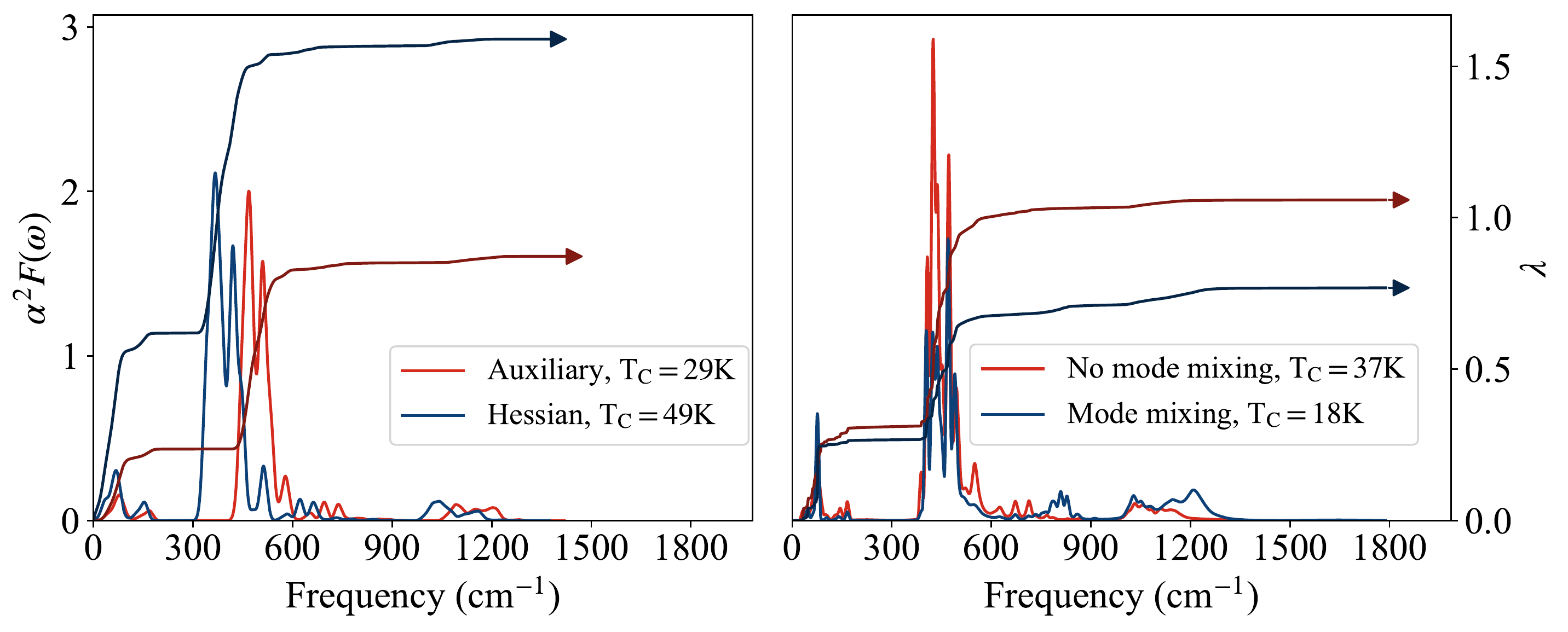}
	\caption{Eliashberg spectral function $\alpha^2F(\omega)$ and integrated electron-phonon coupling constants $\lambda$ of LuH$_3$ (structure at 6 GPa and 300 K) calculated with different methods.}
	\label{a2fs}
\end{figure}

In Fig.~\ref{a2fs} we are showing the Eliashberg spectral function of LuH$_3$ (structure at 6 GPa and 300 K) calculated with different methods. First, we employ standard Gaussian approximation for phonon spectral functions with auxiliary and hessian phonons calculated with SSCHA. In the second case, we use the phonon spectral functions calculated in dynamical bubble approximation with no mode mixing and mode mixing approaches~\cite{our_hydrogen}. The largest T$_\mathrm{C}$ is obtained with Gaussian approximation for hessian phonons, 49 K. This is considerably smaller than what is reported in Ref.~\cite{Dias}. Interestingly, the critical temperature calculated with more advanced methods, fully incorporating the anharmonicity of phonon modes, is smaller. Additionally, there is a large difference between critical temperatures calculated in mode mixing and no mode mixing approaches, with a more correct, mode mixing, approach giving smaller T$_\mathrm{C}$. All methods show that a large part of electron-phonon coupling comes from octahedral modes.

\newpage
\subsection{Phase diagram of Lu-H}

\begin{figure}[h]
	\includegraphics[width=0.8\linewidth]{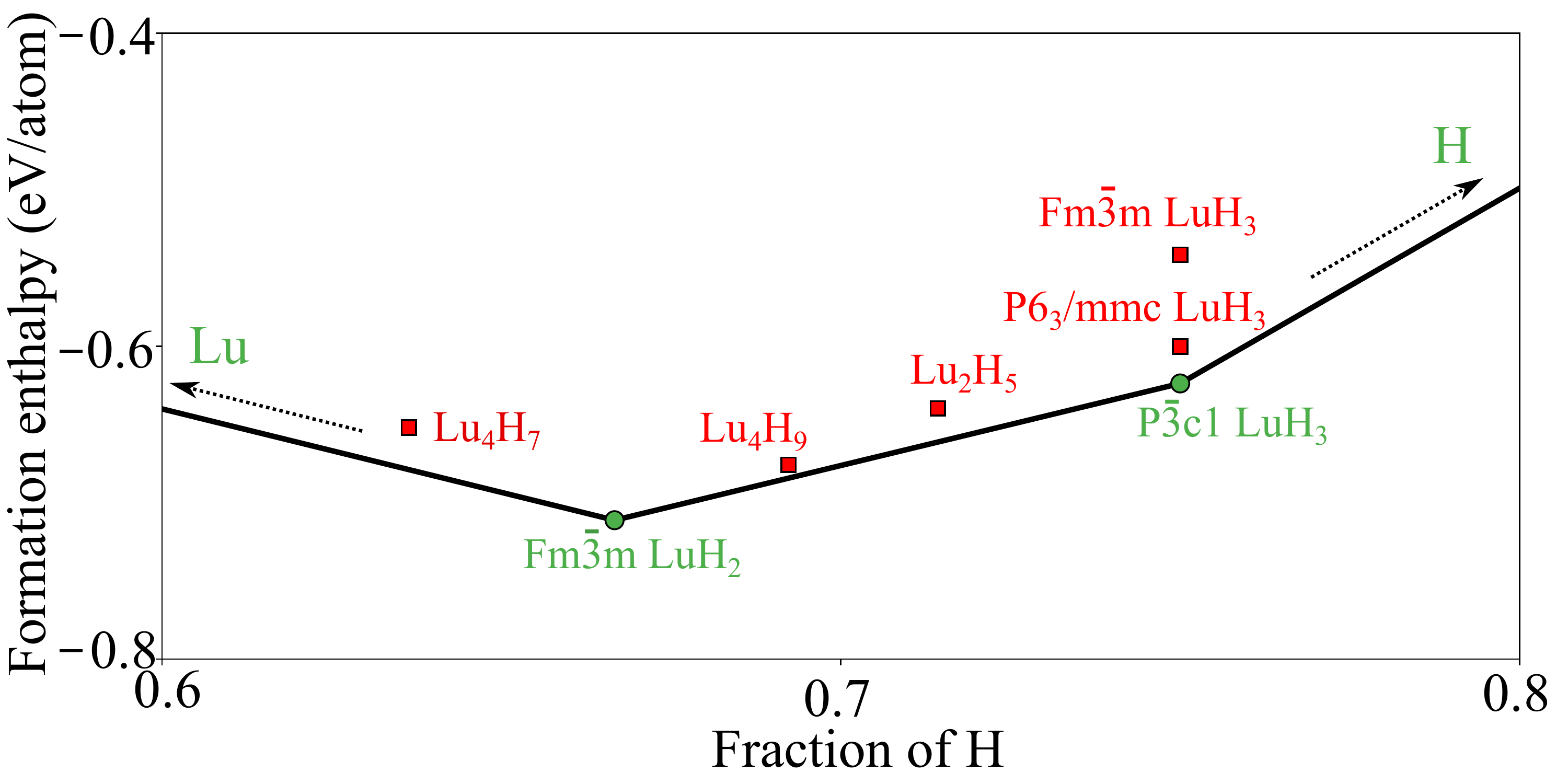}
	\caption{Phase diagram (convex hull) of lutetium hydrides with respect to Lu$_2$ and H$_2$ at 1 GPa and 0 K. Circles are stable phases and the squares denote the phases above the convex hull. The zero-point energy is not accounted for in the calculation of formation enthalpy.}
	\label{convexhull}
\end{figure}

Supplementary Fig.~\ref{convexhull} displays the phase diagram of lutetium hydrides with respect to Lu$_2$ and H$_2$ at 1 GPa. The $Fm\bar{3}m$ LuH$_2$ and $P\bar{3}c1$ LuH$_3$ are stable phases. The $Fm\bar{3}m$ LuH$_3$ is 82 meV/atom above the convex hull.
In the Supplementary Table~\ref{Table:crystalinfo}, we present the crystal structure details of  ${P4/mmm}$ Lu$_2$H$_5$, 
$P\bar{4}3m$ Lu$_4$H$_7$ and  ${Immm}$ Lu$_4$H$_9$ that we have newly predicted. 
To demonstrate the atomic sites of hydrogen atoms, we explicitly show the crystal structure maps of Lu$_2$H$_5$, Lu$_4$H$_7$, and Lu$_4$H$_9$ in Supplementary Fig.~\ref{crystal_map}. 

\begin{longtable}{lllllllllllllllll}
	\caption{Crystal structure details of Lu$_2$H$_5$, Lu$_4$H$_7$ and Lu$_4$H$_9$ at 1 GPa} \\
	Phase & & Cell parameters & & Atom type  & & Wyckoff site & X & Y & Z \\
	\hline
	\hline
	${P4/mmm}$ Lu$_2$H$_5$ &&
	\begin{tabular}[c]{@{}l@{}}$a$ =  3.507~\AA\\
		$b$ =  3.507~\AA\\
		$c$ =  3.507~\AA\\
		$\alpha$ =  90.000$^{\circ}$\\
		$\beta$ =  90.000$^{\circ}$\\
		$\gamma$ =  90.000$^{\circ}$\end{tabular} &&
	\begin{tabular}[c]{@{}l@{}}Lu1\\Lu2\\H1\\H2\end{tabular} &&
	\begin{tabular}[c]{@{}l@{}}1b\\1c\\4i\\1d\end{tabular} &
	\begin{tabular}[c]{@{}l@{}}0.000\\0.500\\0.000\\0.500\end{tabular} &
	\begin{tabular}[c]{@{}l@{}}0.000\\0.500\\0.500\\0.500\end{tabular} &
	\begin{tabular}[c]{@{}l@{}}0.500\\0.000\\0.238\\0.500\end{tabular} &
	\\
	\hline
	$P\bar{4}3m$ Lu$_4$H$_7$ &&
	\begin{tabular}[c]{@{}l@{}}$a$ =  4.974~\AA\\
		$b$ =  4.974~\AA\\
		$c$ =  4.974~\AA\\
		$\alpha$ =  90.000$^{\circ}$\\
		$\beta$ = 90.000$^{\circ}$\\
		$\gamma$ =  90.000$^{\circ}$\end{tabular} &&
	\begin{tabular}[c]{@{}l@{}}Lu1\\H1\\H2\\H3\end{tabular} &&
	\begin{tabular}[c]{@{}l@{}}4e\\3d\\3c\\1a\end{tabular} &
	\begin{tabular}[c]{@{}l@{}}0.751\\0.500\\0.000\\0.000\end{tabular} &
	\begin{tabular}[c]{@{}l@{}}0.751\\0.000\\0.500\\0.000\end{tabular} &
	\begin{tabular}[c]{@{}l@{}}0.751\\0.000\\0.500\\0.000\end{tabular} &
	\\
	\hline
	${Immm}$ Lu$_4$H$_9$ &&
	\begin{tabular}[c]{@{}l@{}}$a$ =  4.885~\AA\\
		$b$ =  4.996~\AA\\
		$c$ =  9.926~\AA\\
		$\alpha$ =  90.000$^{\circ}$\\
		$\beta$ = 90.000$^{\circ}$\\
		$\gamma$ =  90.000$^{\circ}$\end{tabular} &&
	\begin{tabular}[c]{@{}l@{}}Lu1\\Lu2\\Lu3\\H1\\H2\end{tabular} &&
	\begin{tabular}[c]{@{}l@{}}4j\\2a\\2c\\16o\\2b\end{tabular} &
	\begin{tabular}[c]{@{}l@{}}0.500\\0.000\\0.500\\0.744\\0.000\end{tabular} &
	\begin{tabular}[c]{@{}l@{}}0.000\\0.000\\0.500\\0.757\\0.500\end{tabular} &
	\begin{tabular}[c]{@{}l@{}}0.752\\0.000\\0.000\\0.373\\0.500\end{tabular} &
	\\
	
	\hline
	\hline
	\label{Table:crystalinfo}
\end{longtable}

\begin{figure}[h]
	\includegraphics[width=0.8\linewidth]{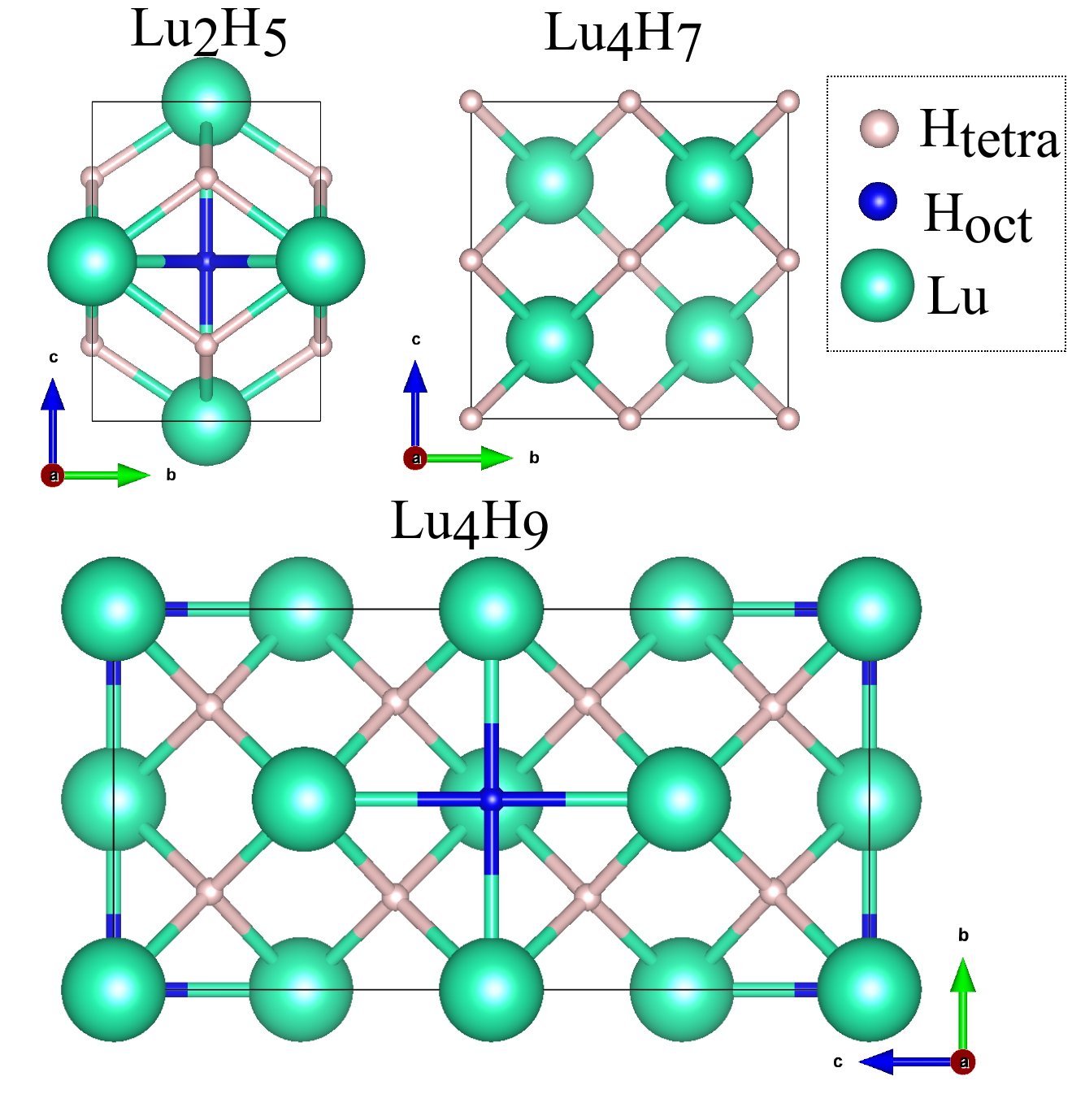}
	\caption{The predicted crystal structures of ${P4/mmm}$ Lu$_2$H$_5$, 
		$P\bar{4}3m$ Lu$_4$H$_7$ and  ${Immm}$ Lu$_4$H$_9$ in our study. H$_{\rm tetra}$ and H$_{\rm oct}$ refer to the hydrogen atoms at the tetrahedral and octahedral sites, respectively. }
	\label{crystal_map}
\end{figure}

\newpage

\end{document}